\documentclass[preprint,onecolumn,nofootinbib]{revtex4}
\pdfoutput=1
\usepackage[colorlinks=true,linkcolor=blue,urlcolor=blue,filecolor=black,citecolor=red,pdfstartview=FitV,pdftitle={},pdfsubject={},pdfkeywords={},pdfpagemode=None,bookmarksopen=true]{hyperref}
\usepackage{graphicx}
\usepackage{amsmath}
\usepackage{amsfonts}
\usepackage{amssymb,ulem}
\usepackage{color,xcolor}%
\usepackage{CJK}
\usepackage{subfigure}

\usepackage{dcolumn}
\newcommand{\blue}{\textcolor[rgb]{0.33,0.33,1.00}}

\begin{document}
\title{A novel holographic quantum phase transition and butterfly velocity}
\author{Guoyang Fu$^{1}$}
\thanks{FuguoyangEDU@163.com}
\author{Xi-Jing Wang$^{1}$}
\thanks{xijingwang@yzu.edu.cn}
\author{Peng Liu $^{2}$}
\thanks{phylp@email.jnu.edu.cn}
\author{Dan Zhang $^{3}$}
\thanks{danzhanglnk@163.com}
\author{Xiao-Mei Kuang$^{1}$}
\thanks{xmeikuang@yzu.edu.cn}
\author{Jian-Pin Wu$^{1}$}
\thanks{jianpinwu@yzu.edu.cn}
\affiliation{
  $^1$ Center for Gravitation and Cosmology, College of Physical Science and Technology, Yangzhou University, Yangzhou 225009, China }
\affiliation{$^2$ Department of Physics and Siyuan Laboratory, Jinan University, Guangzhou 510632, P.R. China}
\affiliation{ $^3$ Key Laboratory of Low Dimensional Quantum Structures and Quantum Control of Ministry of Education, Synergetic Innovation Center for Quantum Effects and Applications, and Department of Physics, Hunan Normal University, Changsha, Hunan 410081, China}

\begin{abstract}
\baselineskip=0.6cm

In this paper, we make a systematical and in-depth exploration on the phase structure and the behaviors of butterfly velocity in an Einstein-Maxwell-dilaton-axions (EMDA) model. Depending on the model parameter, there are two kinds of mechanisms driving quantum phase transition (QPT) in this model. One is the infrared (IR) geometry to be renormalization group (RG) unstable, and the other is the strength of lattice deformation leading to some kind of bifurcating solution.
We also find a novel QPT in the metal phases.
The study on the behavior of the butterfly velocity crossing QPT indicates that the butterfly velocity or its first derivative exhibiting local extreme depends on the QPT mechanism. Further, the scaling behaviors of the butterfly velocity in the zero-temperature limit confirm that different phases are controlled by different IR geometries. Therefore, the butterfly velocity is a good probe to QPT and it also provides a possible way to study QPT beyond holography.

\end{abstract}

\maketitle
\tableofcontents

\section{Introduction}
Quantum phase transition (QPT) is an exotic phenomenon in condensed matter theory (CMT) accompanied by the existence of strong correlation between the microscopic degrees of freedom \cite{Sachdev:2000}. The conventional theoretical tools are usually inadequate to deal with these strongly correlated systems.
The holographic duality offers a powerful method to explore the strongly correlated systems by mapping them into the weakly coupled classical gravity theories \cite{Maldacena:1997re,Gubser:1998bc,Witten:1998qj,Aharony:1999ti}. The holographic duality can shed light on the basic principle hidden in the strongly correlated phenomenon.

Metal-insulator transition (MIT), as an example of QPT, has been widely studied from holography in recent years \cite{Donos:2012js,Donos:2014oha,Donos:2013eha,Donos:2014uba,Ling:2014saa,Baggioli:2014roa,Kiritsis:2015oxa,Ling:2015epa,Ling:2015exa,Ling:2016dck,Mefford:2014gia,Baggioli:2016oju,Andrade:2017ghg,Bi:2021maw}. The main mechanism of holographic MIT is that the lattice operator breaking the translational symmetry induces the infrared (IR) instability, which results in a new IR fixed point corresponding to insulating phase \cite{Donos:2012js,Donos:2014uba}. Different lattice structures give rise to diverse insulating phases in the dual boundary theory. The well-known examples include holographic Peirels insulator, holographic Mott insulator and some novel holographic insulating phases. These holographic insulating models exhibit some appealing characteristics, for instance, the hard gap \cite{Kiritsis:2015oxa,Ling:2015epa,Ling:2016dck} and the commensurability \cite{Andrade:2017ghg} in Mott insulator, the pinned collective mode and gapped single-particle excitation in Peirels insulator \cite{Ling:2014saa}, which are very similar to those found in condensed matter physics.

An important weapon to study QPT is to identify the characteristic quantity that can reflect the properties of QPT.
The butterfly effect could provide an important tool to handle and understand QPT in holographic framework. It has been shown that the butterfly effect ubiquitously exists in holographic systems and hence has become an important characteristic quantity of holographic system \cite{Maldacena:2015waa,Blake:2016wvh,Blake:2016sud,Lucas:2016yfl}. More importantly, it has been found that the butterfly velocity can signalize the quantum and thermal phase transitions \cite{Ling:2016ibq,Ling:2016wuy,Baggioli:2018afg,Liu:2021stu}.

In this paper, we shall study the phase structure and the butterfly effect over an Einstein-Maxwell-dilaton-axions (EMDA) theory \cite{Donos:2014uba}, which is a simple extension to the Q-lattice model studied in \cite{Donos:2013eha}. The IR fixed points of insulating phases in the original holographic Q-lattice model are currently obscure \cite{Donos:2013eha}. But in EMDA theory \cite{Donos:2014uba}, the IR geometry of the insulating phases can be analytically worked out. It can help us further explore the basic principle behind the phenomena.
In addition, the EMDA theory in \cite{Donos:2014uba} introduces a coupling between the lattice and the Maxwell field such that novel ground state solutions are found. Depending on the model parameters, this holographic system exhibits insulating or metallic behavior. In this paper, we further implement a comprehensive and systematic study on the phase structure. In particular, we mainly focus on the phase structure of the strength of the lattice and the lattice wave number for given coupling parameter. We find that the system exhibits not only the original MIT but also a novel quantum phase transition beyond MIT. Also we explore the behaviors of the butterfly velocity crossing QPT and its scaling behaviors in the zero-temperature limit.

The paper is organized as follows. In section \ref{sec-EMDA-model}, we introduce the holographic setup, work out the background solutions and discuss allowed values for the parameter by the IR analysis. In section \ref{PhaseDiagram}, we compute DC conductivity and show the phase diagram to study the properties of these phases. In section \ref{DiagnoseQPT}, we calculate the butterfly velocity and study the relation between butterfly velocity and quantum critical points (QCPs). In section \ref{scalingvb}, we analyze the scaling behaviors of butterfly velocity with temperature for different phases. The conclusions and discussions are presented in section \ref{conclusion}. In Appendix \ref{appendix}, we give a general form of the EMDA action and give the corresponding equations of motion.

\section{Holographic Background}\label{sec-EMDA-model}

In this paper, we consider the following EMDA action \cite{Donos:2014uba}
\begin{eqnarray}\label{EMDA-Action}
S= {}\int d^{4}x \sqrt{-g} \left[ R +6 \cosh\psi - \frac{3}{2} [ (\partial \psi)^2+4\sinh^2\psi (\partial{\chi})^2 ] - \frac{1}{4} \cosh^{\gamma /3}(3\psi)F^2 \right]\,.
\end{eqnarray}
$F=dA$ is the field strength of the Maxwell field $A$. $\psi$ is a neutral scalar field dubbed dilaton field.
$\chi$ is the axion field. Here, we assume that $\chi$ only depends on one of the two spatial directions of the dual field theory, which leads the background to be anisotropic.
This model can also be constructed by involving a dilatonic coupling in the linear axion model \cite{Andrade:2013gsa}.

We take the following anisotropic background ansatz
\begin{eqnarray} \label{metric}
ds^2&&=\frac{1}{z^2}\left[-(1-z)p(z)U(z)dt^2+\frac{dz^2}{(1-z)p(z)U(z)}+V_1(z) dx^2+V_2(z) dy^2\right], \\ \nonumber
A&&=\mu(1-z)a(z) dt,  \\ \nonumber
\psi&&=z^{3-\Delta}\phi(z), \\ \nonumber
\chi&&=\hat{k} x,
\end{eqnarray}
where $p(z)=1+z+z^2-\mu^2 z^3/4$ and the conformal dimension of the scalar field $\psi$ is $\Delta=2$. In our convention, the black hole horizon and AdS boundary are located at $z=1$ and $z=0$ respectively.

From the action \eqref{EMDA-Action}, we obtain four second order  ordinary differential equations (ODEs) for $V_1\,, V_2\,, a\,, \phi$ and one first order ODE for $U$. To solve those ODEs numerically, we impose the boundary conditions on the conformal boundary
	\begin{eqnarray}
		U(0)=1\,, \ V_1(0)=1\,, \ V_2(0)=1\,, \ a(0)=1\,, \ \phi(0)=\hat{\lambda}\,,
	\end{eqnarray}
	where the $\hat{\lambda}$ is the source of the scalar field operator in the dual field theory and characterize the lattice deformation in this theory.
	The above boundary conditions come from the requirement of the asymptotic AdS$_4$ on the conformal boundary. At the horizon ($z=1$), we then impose the regular boundary conditions. Further, the Hawking temperature can be given as 
	\begin{eqnarray}
		\hat{T}=\frac{(12-\mu^2)U(1)}{16 \pi}\,.
	\end{eqnarray} 
	We focus on the canonical ensemble by setting the chemical potential $\mu$ as the scaling unit. Thus, after fixing the parameter $\gamma$, this system can be depicted by the three dimensionless parameters $\{T , \lambda , k \} \equiv \{\hat{T}/\mu , \hat{\lambda}/\mu , \hat{k}/\mu \}$.

When $\chi=\psi=0$, the model \eqref{EMDA-Action} admits the RN-AdS black hole solution. In the zero-temperature limit, the IR geometry is AdS$_2 \times \mathbb{R}^2$, which is given by
\begin{eqnarray} 
&&
ds^2=-6\zeta^2 dt^2+\frac{d\zeta^2}{6\zeta^2}+dx^2+dy^2\,,
\nonumber
\\
&&
A_t= 2\sqrt{3}\zeta\,.
\label{AdS2_R2}
\end{eqnarray}
It is helpful to study the perturbations about the IR fixed point AdS$_2 \times \mathbb{R}^2$, which present a preliminary picture about this model and also guide the numerical exploration. To this end, we consider the following perturbations,
\begin{eqnarray} \label{scalar_pertubation}
&&g_{tt}=g_{\zeta\zeta}^{-1}=6\zeta^2(1+u_1 \zeta^\delta)\,, \ \ g_{xx}=e^{2V_1}\,, \ \ g_{yy}=e^{2V_2}\,, \ \ a=2\sqrt{3}\zeta (1+a_1 \zeta^\delta)\,,  \nonumber \\
&&V_1=v_{10}(1+v_{11}\zeta^\delta)\,, \ \ V_2=v_{20}(1+v_{21}\zeta^\delta)\,, \ \ \delta\psi=\psi_0 \zeta^\delta\,,\ \  \  \ \chi=kx\,, \ \
\end{eqnarray}
where $u_1\,,a_1\,,v_{10}\,,v_{11}\,,v_{20}\,,v_{21}$ are small constants, and $\delta$ is the scaling dimension of the deformation.
Substituting Eq.\eqref{scalar_pertubation} into the equations of motion, we can work out the scaling dimension of the scalar field operator $\delta^\psi$ as
\begin{eqnarray} \label{delta_psi}
\delta_{+}^{\psi}=-\frac{1}{2}+\frac{1}{6}\sqrt{24 e^{-2 v_{10}} k^2-3(12\gamma+1)}.
\end{eqnarray}

It is easy to see that if the relation
\begin{eqnarray}
\label{relation-gamma-k}
2e^{-2 v_{10}}k^2\geq 1+3\gamma\,,
\end{eqnarray}
is satisfied, we always have $\delta_{+}^{\psi}\geq 0$, which means that the IR solution is always RG stable. Especially, we notice that when $k=0$, the mode $\delta_{+}^{\psi}$ is minimized. Based on the above observation, we categorize this system into the following three cases:
\begin{itemize}
  \item Case I: For $-1<\gamma \leq -1/3$, we always have $\delta_{+}^{\psi}>0$ at $k\neq 0$, which corresponds to an irrelevant deformation in IR. Notice that the lower bound $\gamma>-1$ is set by the requirement for the fixed point solutions to exist. See Section 2 in \cite{Donos:2014uba} for detailed discussion.
  \item Case II: For $-1/3<\gamma \leq -1/12$, we find that $\delta_{+}^{\psi}<0$ at $k=0$. It means the IR solution to be RG unstable. We notice that at $k\neq 0$, the IR solution is also RG unstable when $2e^{-2v_{10}}k^2< 1+3\gamma$ as the case of $k=0$, but holds stable if the relation \eqref{relation-gamma-k} is not violated. Therefore, when reducing $k$ or increasing $\lambda$, the IR solution is RG unstable, which induces a MIT \cite{Donos:2014uba}.
  \item Case III: When $\gamma>-1/12$, $\delta_{+}^{\psi}$ becomes complex at $k=0$ such that the BF bound is violated resulting in a dynamical instability. Therefore, the system develops into a novel black hole with scalar hair. Depending on the parameter $\gamma$, this novel black hole with scalar hair has different ground states at zero temperature \cite{Donos:2014uba}. By the DC and AC conductivities over the IR fixed point, we can determine that the ground state of this black hole is insulating for $-1/12<\gamma<3$, and metallic for $\gamma>3$ \cite{Donos:2014uba}.
\end{itemize}

\section{phase diagram}\label{PhaseDiagram}

Ref.\cite{Donos:2014uba} have constructed some specific black hole solutions for certain values of $\gamma$. Such black holes at zero temperature limit correspond to new IR fixed points. But for the given $\gamma$ and the temperature $T$, the full phase diagram over $\lambda$ and $k$ is absent. In this section, we make a detailed exploration on the full phase diagram as $\lambda$ and $k$ are varied.

At extremely low temperatures, the metallic phase and insulating phase can be distinguished by the change of the $\sigma_{DC}$ with the temperature. Specifically, the metallic phase satisfies $\partial_T\sigma_{DC}<0$ and the insulating phase satisfies $\partial_T\sigma_{DC}>0$.
The critical points can be identified as those points where $\partial_T\sigma_{DC}=0$.

For a class of EMDA theory \eqref{appendix_action} with ansatz \eqref{appendix_metric}, one can calculate the DC conductivity analytically by using the horizon data via the membrane paradigm \cite{Iqbal:2008by,Donos:2014uba,Donos:2014cya}
\begin{eqnarray}\label{exp_sigmadc1}
\sigma_{DC}=\sqrt{C_1 C_2} Z(\psi )\left(\frac{1}{C_1}+\frac{a'^2 Z(\psi )}{c B D k^2 Y(\psi )}\right)\Bigg{|}_{z\to 1},
\end{eqnarray}
where the prime denotes derivative with respect to bulk radial direction.
Here one can combine the specific model \eqref{EMDA-Action} with ansatz \eqref{metric} and the above formula \eqref{exp_sigmadc1} to obtain a concrete expression for DC conductivity.

\begin{figure}[ht!]
	\centering
	\includegraphics[width=0.45\textwidth]{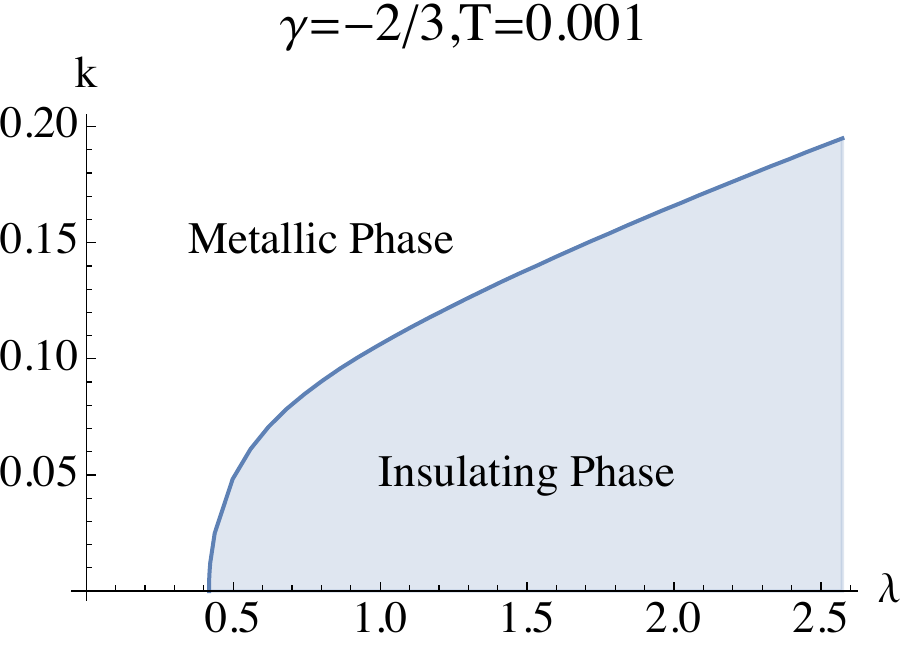}\hspace{2mm}
	\includegraphics[width=0.45\textwidth]{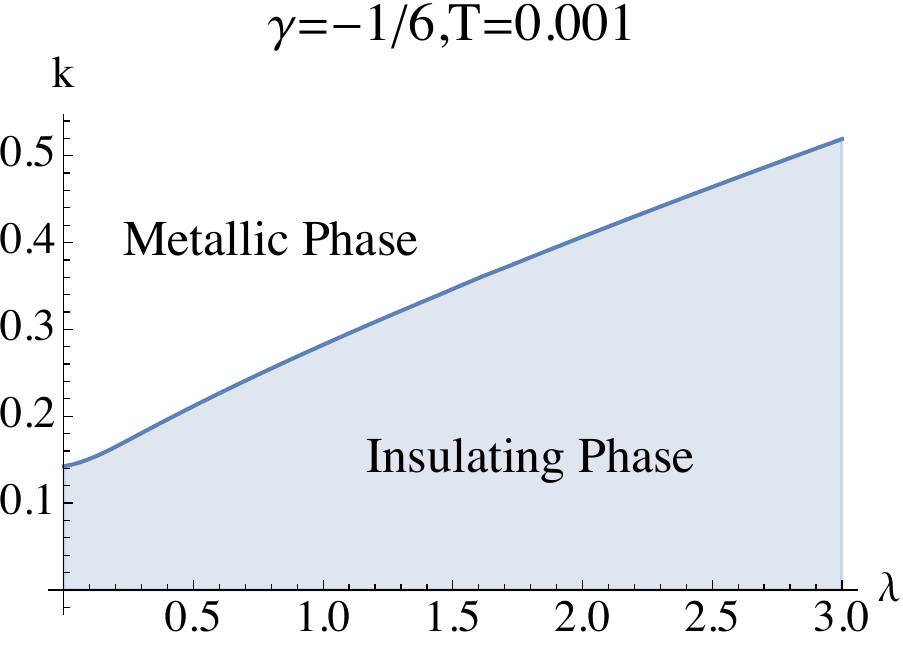}\vspace{0.4mm}\ \\
	\includegraphics[width=0.45\textwidth]{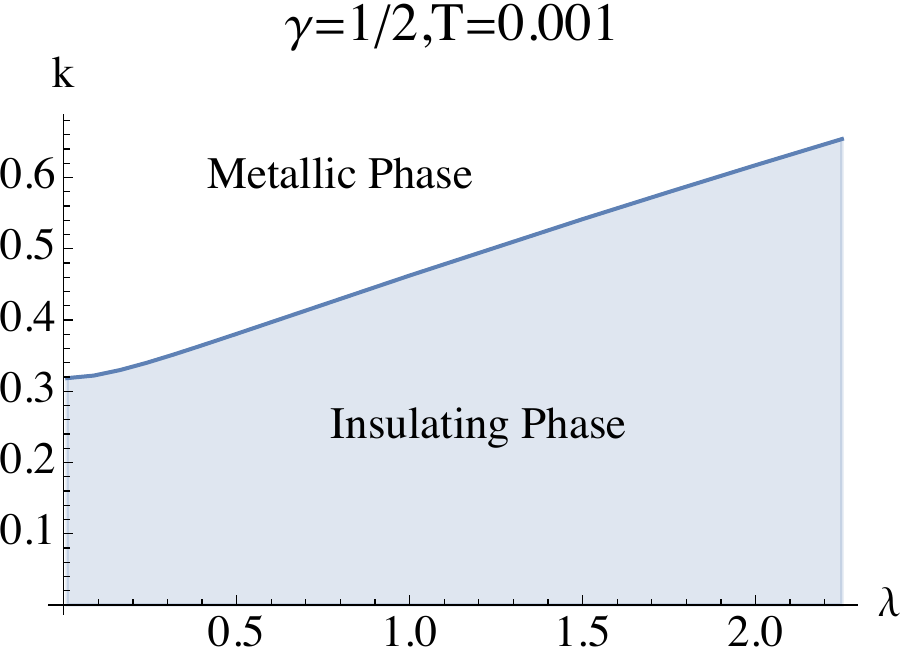}\hspace{2mm}
	\includegraphics[width=0.45\textwidth]{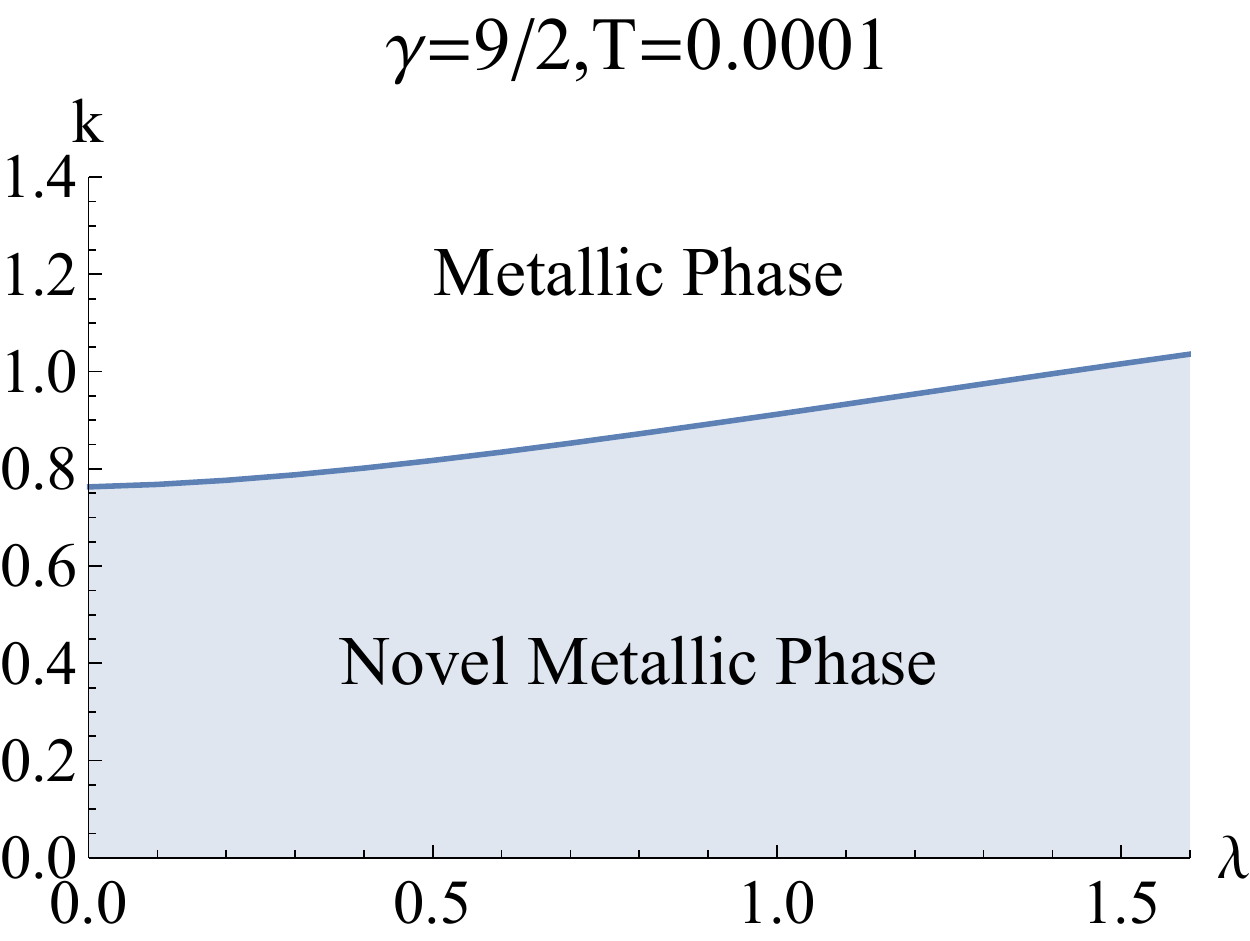}\hspace{0.4mm}
	\caption{The phase diagram over the $\{\lambda, k\}$ with different $\gamma$ at some fixed temperatures. The blue line is the critical line of the phase transition.}
	\label{phase diagram}
\end{figure}

Based on the above discussion about the region of the parameter $\gamma$, the phase diagrams over $(\lambda, k)$ with some specific values of $\gamma$ at a low temperatures\footnote{For $\gamma=-2/3$, $-1/6$ and $1/2$, we set the temperature as $T=0.001$ and for $\gamma=9/2$, $T=0.0001$. We have checked that further decreasing the temperature will not lead to the phase diagram to be changed substantially.}, have been shown in Fig.\ref{phase diagram}. It is obvious that in the lower right case, a novel metallic phase emerges in small $k$ rather than the insulating phase in other cases. We will carefully analyze the physics in each phase diagram as follows.

The upper left plot in Fig.\ref{phase diagram} shows the phase diagram for $\gamma=-2/3$.
We observe that when the strength of the lattice $\lambda$ is small, the phase is metallic even for small $k$.
It is because the IR geometry at $T=0$ is an AdS$_2 \times \mathbb{R}^2$ as illustrated in the IR analysis above.
It was confirmed by the fact that the black hole has non-zero entropy density at zero temperature (see Fig.1(a) in \cite{Donos:2014uba}).
Increasing $\lambda$ with fixed $k$ we find that the MIT emerges. Indeed, when $\lambda$ increases, the black hole in the limit of $T=0$ flows to a new IR fixed point.
It was also illustrated by the black hole entropy which vanishes at zero temperature (see Fig.1(a) in \cite{Donos:2014uba}).
The study of the butterfly effect below also confirms this point. From the observation above, we conclude that the strength of the lattice deformation $\lambda$ drives an MIT.
However, since the IR geometry in this case is RG stable as analyzed above, the mechanism driving the MIT must be distinct from the stability analysis. A possible understanding of this novel mechanism is the existence of bifurcating solutions as argued in \cite{Donos:2012js}.

Different from the case I, the MIT always exists for any $\lambda$ for $\gamma=-1/6$ (the upper right plot in Fig.\ref{phase diagram}).
For the region of larger $k$ and smaller $\lambda$, the lattice deformation is an irrelevant deformation and the IR geometry is still the AdS$_2 \times \mathbb{R}^2$ which corresponds to the metallic phase.
When increasing the strength of the lattice deformation $\lambda$ or reducing wave vector $k$, the lattice deformation becomes RG relevant inducing a transition from the AdS$_2 \times \mathbb{R}^2$ to a new IR fixed point, and thus an MIT happens.
This mechanism is proposed in \cite{Donos:2012js}. It is also the MIT mechanism of the original Q-lattice models studied in \cite{Donos:2013eha}.

For the case of $\gamma>-1/12$, the system develops into a novel black hole with scalar hair, which has different ground states at zero temperature depending on the parameter $\gamma$. The different ground states exhibit very different phase structures. The phase diagrams of two representative examples of $\gamma=1/2$ and $\gamma=9/2$ are shown below in Fig.\ref{phase diagram}. For $\gamma=1/2$, the ground state is insulating. The phase diagram is very similar to the case of $\gamma=-1/6$ (the below left in Fig.\ref{phase diagram}).

\begin{figure}[ht!]
	\centering
	\includegraphics[width=0.45\textwidth]{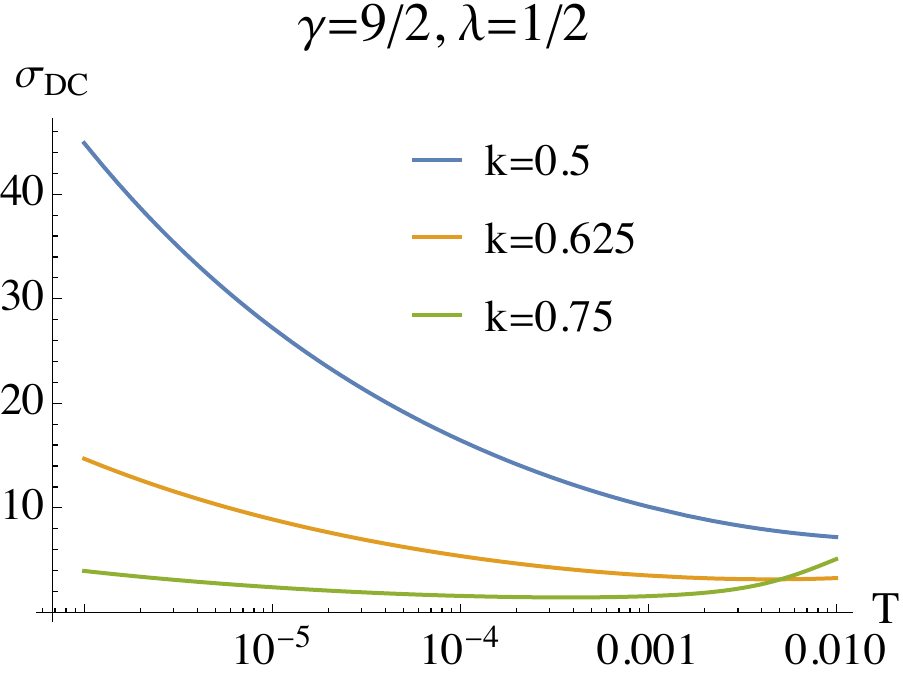}\hspace{2mm}
	\includegraphics[width=0.45\textwidth]{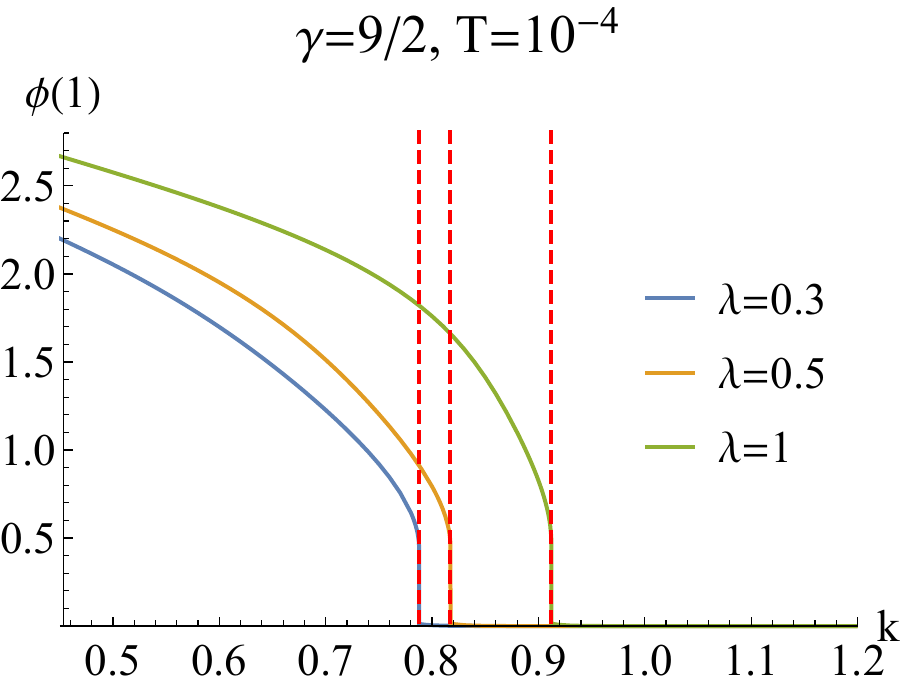}\hspace{0.4mm}
	\caption{Left plot: The DC conductivity as the function of $T$ for $\gamma=9/2,\,\lambda = 1/2$ at different values of $k$. Right plot: The value of the dilaton field $\phi$ at the horizon. The red dashed lines are the positions of QCPs for specific $\lambda$.}
	\label{phase diagram 2}
\end{figure}

When $\gamma=9/2$, the ground state is metallic. The story is completely different. The left plot in Fig.\ref{phase diagram 2} shows the DC conductivity vs the temperature $T$, from which we see that for small $k$, the system is indeed metallic at extremely low temperatures. This is different from the cases studied above and the usual Q-lattice model \cite{Donos:2013eha}, in which the system is insulating in the region of small $k$ \cite{Ling:2015dma}. Next, we show that the IR geometry of this metallic phase is a hyperscaling violation geometry. Therefore, we conclude that we obtained a novel metallic phase with non-AdS$_2 \times \mathbb{R}^2$ IR geometry\footnote{In our previous work of two-dimensional Q-latticed holographic systems \cite{Liu:2021stu}, we have found an anisotropic metallic phase with non-AdS$_2 \times \mathbb{R}^2$ IR geometry.}.

Further increasing $k$ at fixed values of $\lambda$ to exceed some critical values, we find a phase transition from the novel metallic phase with hyperscaling violation IR geometry to the normal metallic phase with AdS$_2 \times \mathbb{R}^2$ IR geometry. We plot the value of the dilaton field $\phi(1)$ at the horizon in the right plot in Fig.\ref{phase diagram 2}. As $k$ increases, $\phi(1)$ decreases and finally tends to zero, for which the lattice deformation becomes irrelevant in the IR, leading to an AdS$_2 \times \mathbb{R}^2$ geometry. It is obvious that this phase transition is indeed induced by the transition between different IR fixed points as described in \cite{Donos:2012js}.
To sum up, for the case of $\gamma=9/2$, we have a phase transition from a novel metallic phase with non-AdS$_2 \times \mathbb{R}^2$ IR geometry to a normal metallic phase with AdS$_2 \times \mathbb{R}^2$ IR geometry. The phase diagram is exhibited in the lower right plot in Fig.\ref{phase diagram}.

\section{Diagnosing the QPT by butterfly velocity}\label{DiagnoseQPT}
The previous studies revealed that in holographic models, the first-order derivative of the butterfly velocity with respect to system parameters diagnoses QCP with local extremes in zero-temperature limit \cite{Ling:2016ibq,Liu:2021stu,Baggioli:2018afg}. The EMDA model studied here exhibits richer phase structures, which provide a platform to further test the robustness of the relation between the butterfly velocity and QCP.

The butterfly velocity can be obtained by the shockwave solution near the horizon. The early researchs on the butterfly velocity in holography focused on the isotropic systems \cite{Blake:2016wvh}. Later, the anisotropic system and even the most general case have been investigated \cite{Ling:2016wuy,Blake:2017qgd,Ling:2017jik,Jeong:2017rxg}. Based on the formula derived by \cite{Jeong:2017rxg} and our convention \eqref{metric}, the expression of the butterfly velocity $v_B$ along $x$ direction are given by
\begin{eqnarray}\label{exp_butterfly}
v_B=\sqrt{\frac{-2\pi T \mu V_2}{V_2(V_1'-2V_1)+V_1(V_2'-2V_2)}}\Bigg{|}_{z=1}\,,
\end{eqnarray}
where the prime denotes the derivative with respect to $z$.
For simplicity, we shall fix $\lambda$ and study the dependence of butterfly velocity $v_B$ on $k$. We expect that very similar phenomena can be obtained when varying $\lambda$ at fixed values of $k$. In particular, we mainly focus on the behaviors of the butterfly velocity $v_B$ near QCP in extremal low temperatures.

\begin{sloppypar}
	\begin{figure}[]
		\centering
		\includegraphics[width=0.45\textwidth]{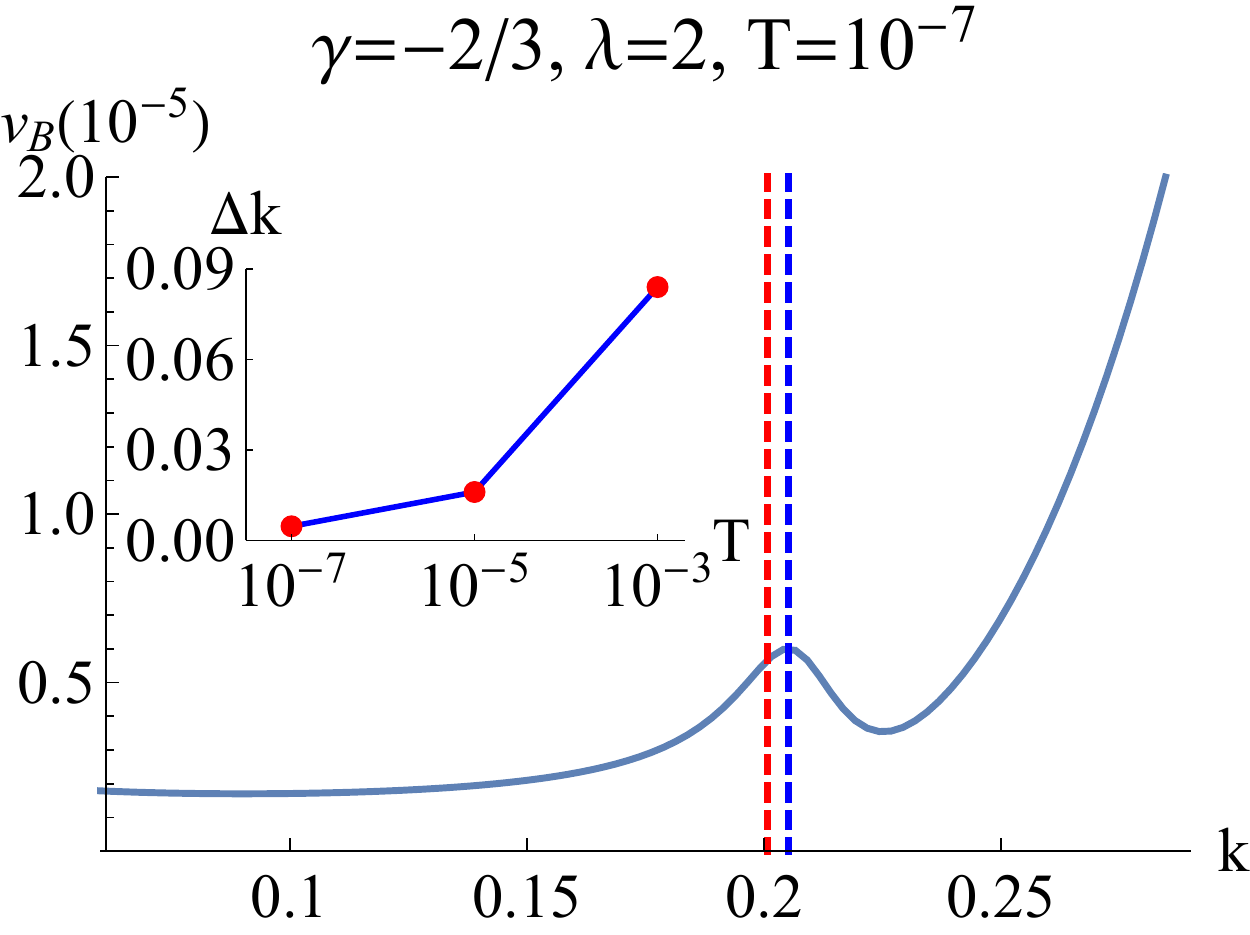}\hspace{5mm}
		\caption{The butterfly velocity $v_B$ as the function of $k$ for $\gamma=-2/3$ at $T=10^{-7}$. The inset plot is for the temperature dependence of $\Delta k$\blue, which is the difference between the position of QCP and the local maxima of $v_B$. The red dashed line is the position of QCP and the blue dashed line is the local maxima of $v_B$. Here we fix $\lambda=2$.}
		\label{butterfly4}
	\end{figure}
\end{sloppypar}

Fig.\ref{butterfly4} shows the butterfly velocity $v_B$ as the function of $k$ for $\gamma=-2/3$ at $T=10^{-7}$, for which the butterfly velocity $v_B$ itself exhibits a local maxima.
Moreover, we also find that the difference $\Delta k$ between the position of QCP and local maxima of $v_B$ decreases monotonically with the decreasing temperature (the inset plot in Fig.\ref{butterfly4}). Therefore, we conclude that in this case, $v_B$ itself can capture the QPT in zero-temperature limit. It is different from the previous result that the first derivative of $v_B$ characterizes QPT in the usual holographic Q-lattice model studied in \cite{Ling:2016ibq,Liu:2021stu}. It is because in this case, the IR geometry is RG stable and the QPT is induced by the strength of the lattice deformation resulting in some kind of bifurcating solution, which is different from the QPT mechanism in \cite{Ling:2016ibq,Liu:2021stu} that the QPT is induced by RG to be unstable.

\begin{sloppypar}
	\begin{figure}[]
		\centering
		\includegraphics[width=0.45\textwidth]{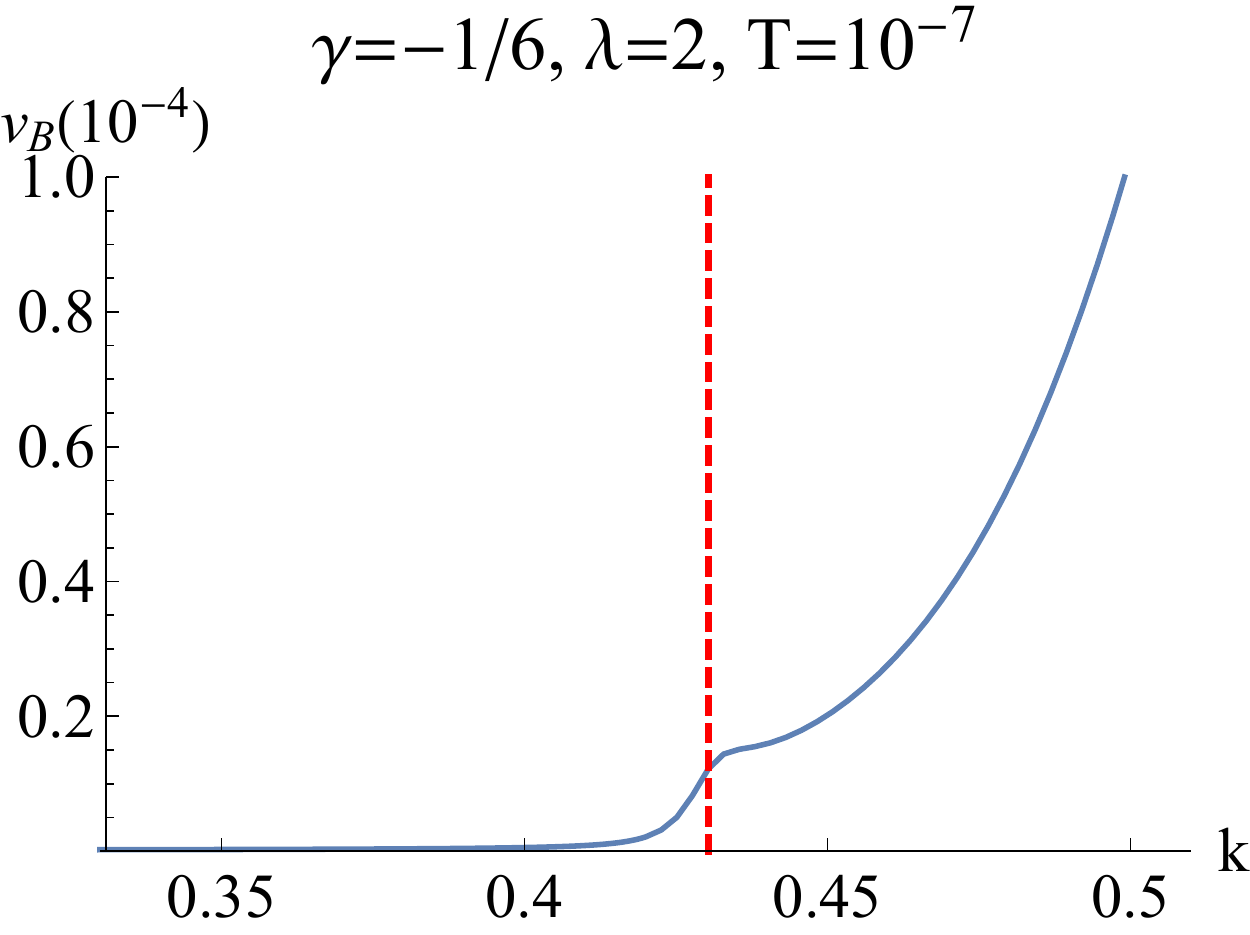}\hspace{5mm}
		\includegraphics[width=0.45\textwidth]{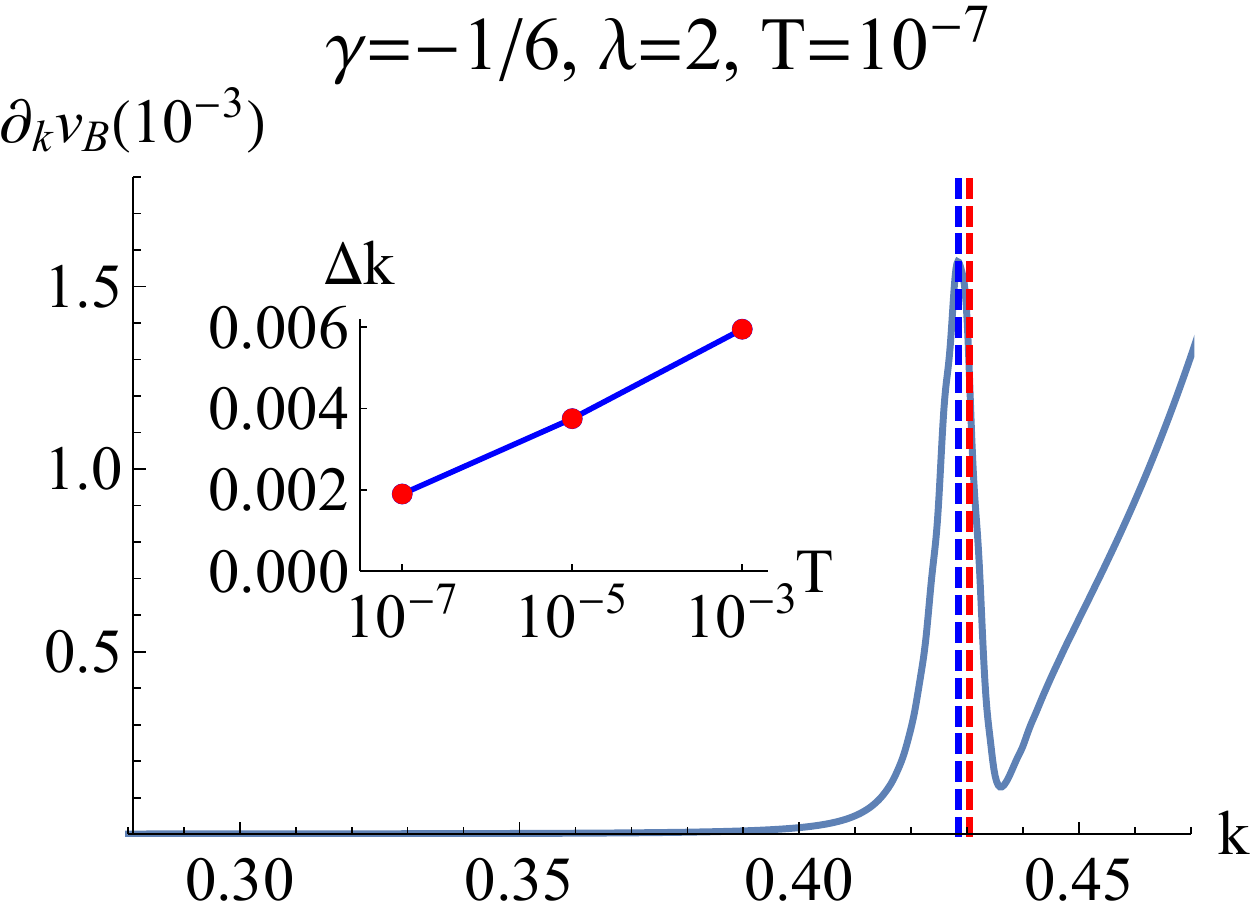}\ \\
		\caption{The butterfly velocity $v_B$ and its derivative with respect to $k$ as the function of $k$ for $\gamma=-1/6$ at $T=10^{-7}$. The inset in the right plot is for the temperature dependence of $\Delta k$, which is the difference between the positions of QCPs and the local maxima of $\partial_k v_B$. The red dashed lines are the positions of QCPs and the blue dashed lines are the local maxima of $\partial_k v_B$. Here we fix $\lambda=2$.}
		\label{butterfly1}
	\end{figure}
\end{sloppypar}

For the case of $\gamma=-1/6$, MIT is induced by the instability of the IR geometry. From Fig.\ref{butterfly1}, we observe that $v_B$ increases with $k$ monotonically, even when the system transits from the insulating phases to metallic phases. Especially, $v_B$ increases by orders of magnitude when transiting from insulating phases to metallic phases. Based on this observation, it is expected to diagnose the QCP by the local extreme of $\partial_k v_B$. We confirm this observation by plotting $\partial_k v_B$ as the function of $k$ in the right plot in Fig.\ref{butterfly1}. Evidently, the position of the local maxima of $\partial_k v_B$ is very close
to the QCP. Further, we use $\Delta k$ to denote the difference between the positions of QCP and local maxima of $\partial_k v_B$ shown in the inset of Fig.\ref{butterfly1}. We find that $\Delta k$ decreases monotonically with the decreasing temperature. Therefore, we conclude that in case II of the EMDA model, the local extreme of $\partial_k v_B$ can diagnose the QPT in the limit of zero temperature. It is consistent with previous discovery in \cite{Ling:2016ibq,Liu:2021stu,Baggioli:2018afg}.

\begin{sloppypar}
 \begin{figure}[]
 \centering
    \includegraphics[width=0.45\textwidth]{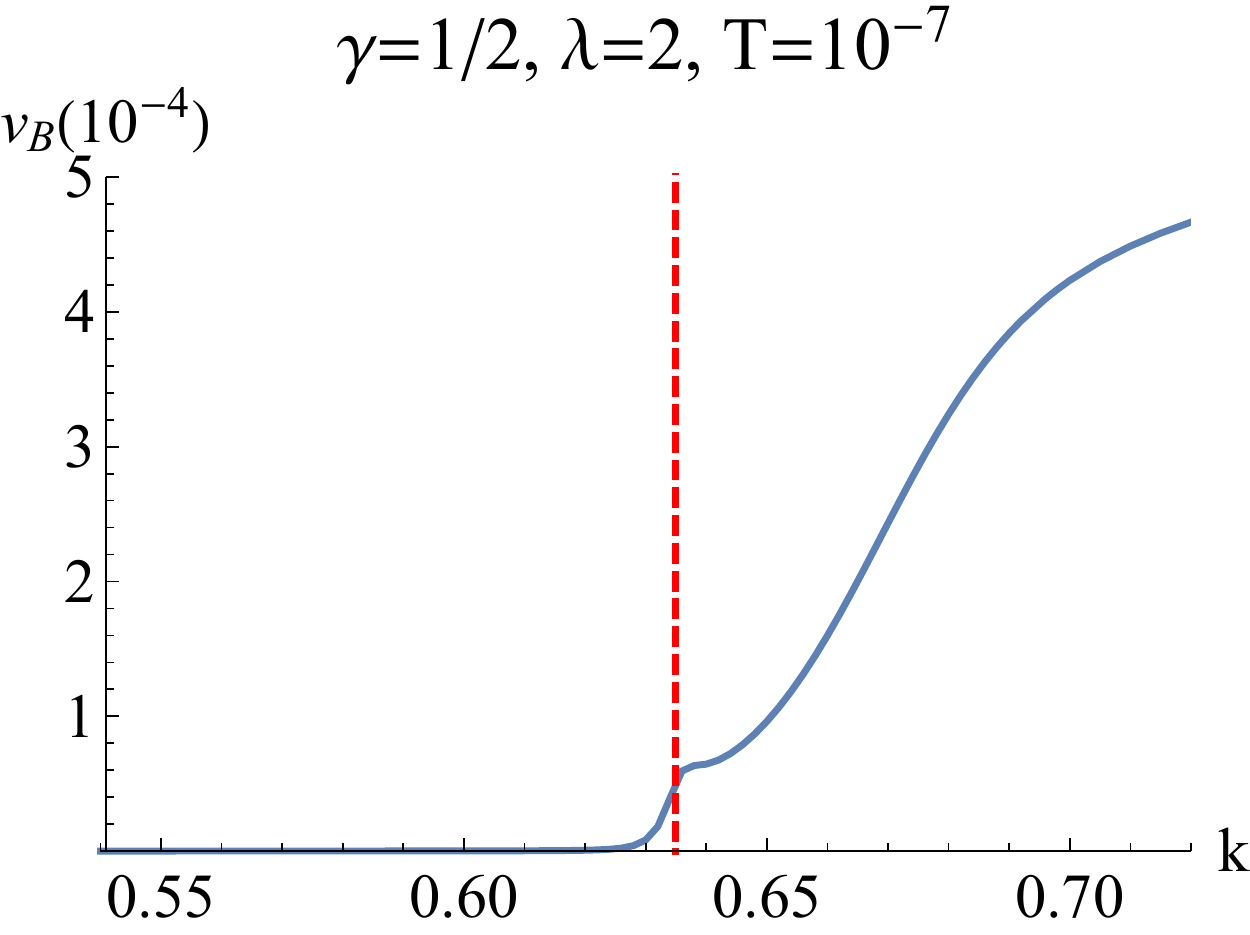}\hspace{5mm}
    \includegraphics[width=0.45\textwidth]{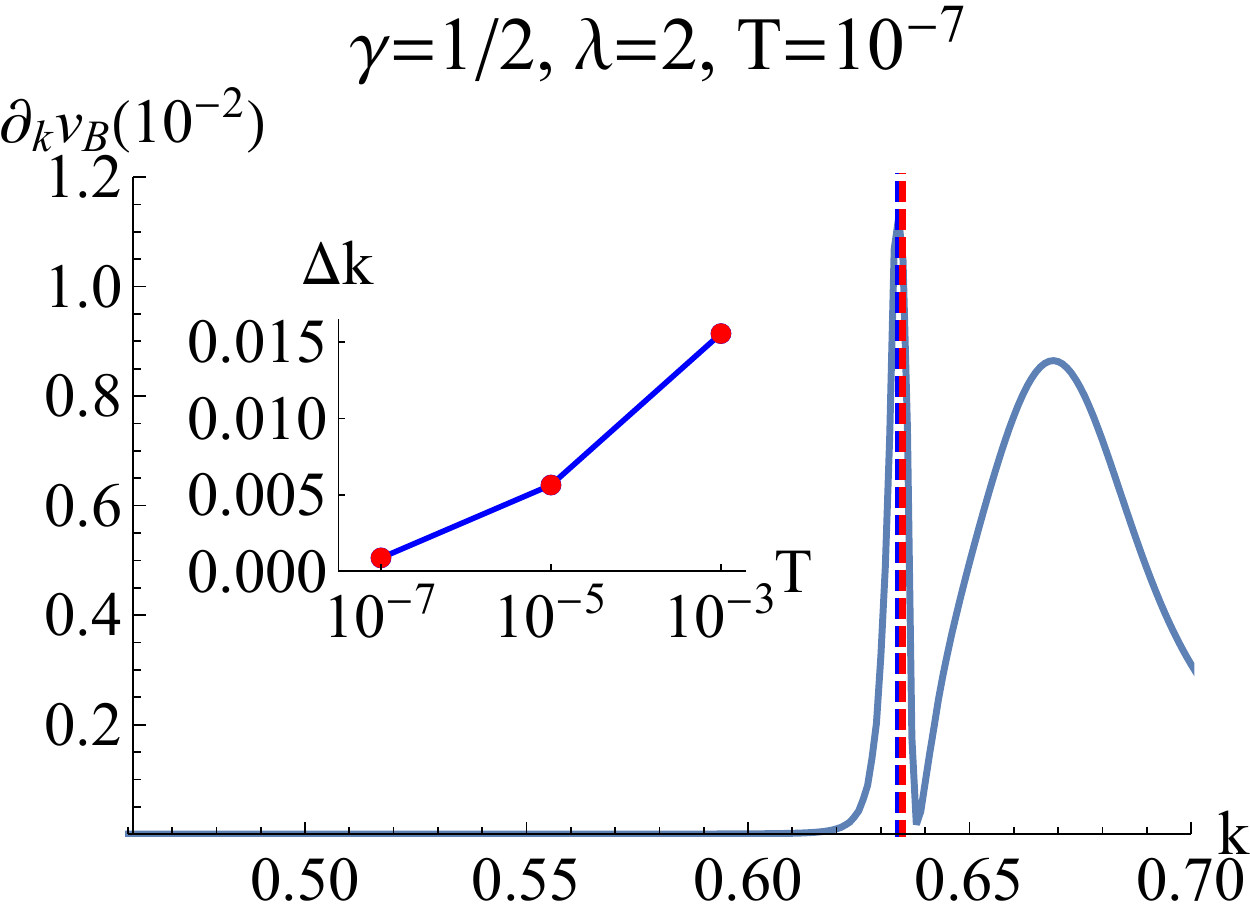}\ \\
  \caption{The butterfly velocity $v_B$ and its derivative with respect to $k$ as the function of $k$ for $\gamma=1/2$ at $T=10^{-7}$. The inset in the right plot is for the temperature dependence of $\Delta k$, which is the difference between the positions of QCPs and the local maxima of $\partial_k v_B$. The red dashed lines are the positions of QCPs and the blue dashed lines are the local maxima of $\partial_k v_B$. Here we fix $\lambda=2$.}
            \label{butterfly2}
\end{figure}
\end{sloppypar}
\begin{figure}[ht!]
    \centering
    \includegraphics[width=0.45\textwidth]{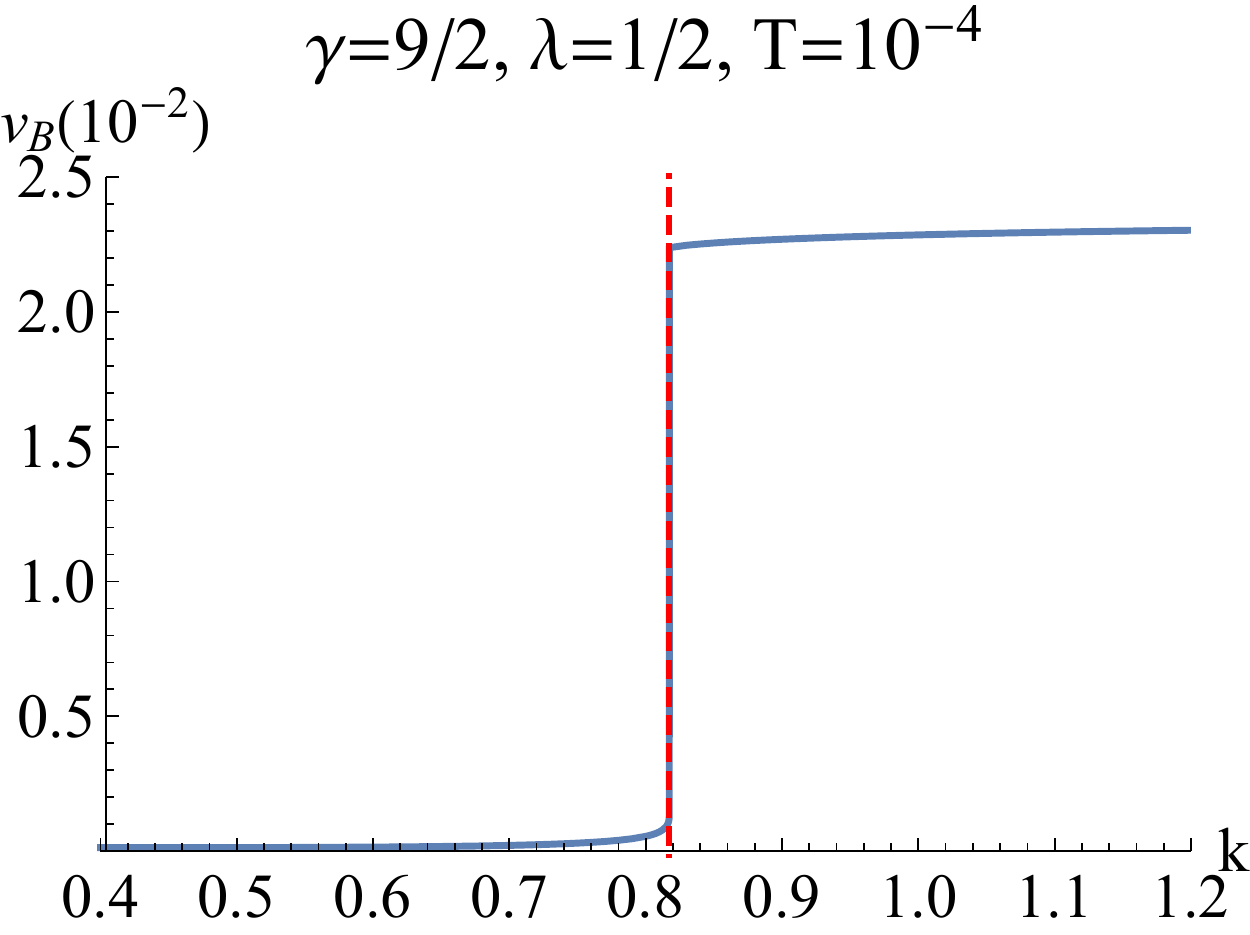}\hspace{5mm}
    \includegraphics[width=0.45\textwidth]{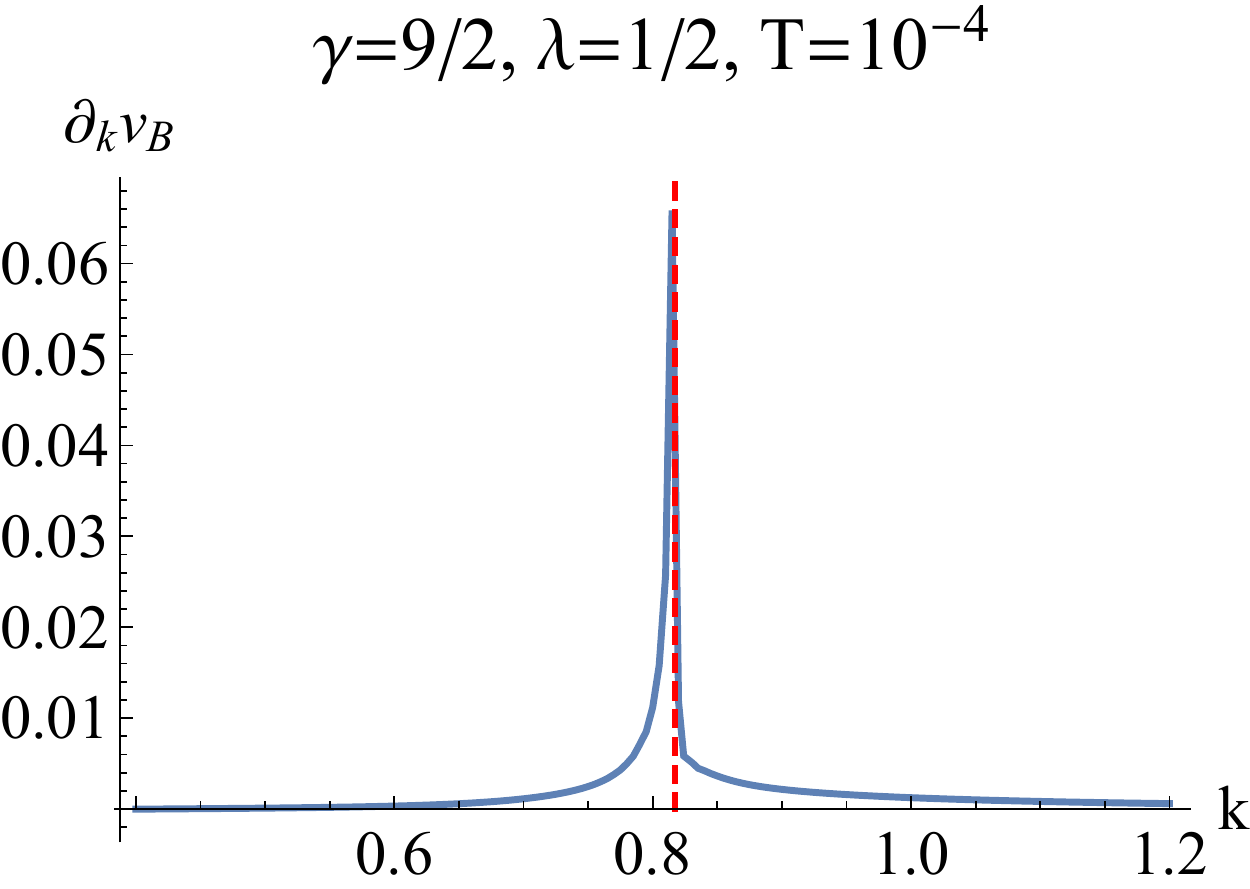}
\caption{The butterfly velocity $v_B$ and its derivative with respect to $k$ as the function of $k$ for $\gamma=9/2$ at $T=10^{-4}$. The red dashed lines are the positions of QCPs. Here we choose $\lambda=1/2$.}
    \label{butterfly_g_9f2}
\end{figure}

Now, we study case III ($\gamma>-1/12$), for which we have a novel black hole with scalar hair. For $\gamma=1/2$, the black hole with scalar hair has an insulating ground state and the phase diagram over the $\{\lambda,k\}$ is similar to that of case II. We find that the behavior of $v_B$ and its derivative are also very similar to that case II as expected (see Fig.\ref{butterfly2}).

For the scalarized black holes corresponding to metallic ground states (representative example for $\gamma=9/2$), we find that increasing $k$ across the critical points, $v_B$ abruptly increases from a small value to a large value (Fig.\ref{butterfly_g_9f2}). Therefore, it is also expected that the local extreme of $\partial_k v_B$ can characterize the QPT in this case. The right plot in Fig.\ref{butterfly_g_9f2}, which shows the behavior of the derivative of $v_B$ with respect to $k$, confirms this expectation.

In summary, we can conjecture that the butterfly velocity itself or its first-order derivative characterizing QPT depends on the mechanisms resulting in QPT. When the IR geometry is RG stable, it is likely the strength of the lattice deformation induces some kind of bifurcating solution resulting in the MIT. For this case, $v_B$ itself characterizes QPT. If MIT is induced by the instability of the IR geometry, QPT can be captured by the first derivative of $v_B$.

\section{The scaling behaviors of the butterfly velocity}\label{scalingvb}

The butterfly velocity itself or its first-order derivative characterizing QPT indicates that the theory flows to different IR fixed points for different phases. We shall further illustrate this issue by exploring the scaling behaviors of the butterfly velocity.

Since in our present EMDA model, the IR geometry in the zero-temperature limit is clearly known, we can analytically work out the scaling behaviors of the butterfly velocity. When the IR geometry is the AdS$_2\times \mathbb{R}^2$, substituting Eq. \eqref{AdS2_R2} into  Eq. \eqref{exp_butterfly}, it is easy to find that $v_B$ follows the behavior of $v_B\sim T^{\alpha}$ with $\alpha=1/2$. When the theory flows to the new IR fixed point with \cite{Donos:2014uba}
\begin{equation}\label{fixed_point_sol}
	g_{tt}=g_{\zeta\zeta}^{-1} \sim \zeta^{u_1}\,, \ \ g_{xx}\sim \zeta^{v_{11}}\,, \ \ g_{yy} \sim \zeta^{v_{22}}\,, \ \ a\sim \zeta^{a_1}\,, \ \ e^{\phi}\sim \zeta^{\phi_1}\,,
\end{equation}
where
\begin{eqnarray}
	&&
	u_1=\frac{2 \left(\gamma ^2+3 \gamma +10\right)}{\gamma ^2+4 \gamma +11}\,,\,\,\,
	v_{11}=-\frac{2(\gamma +1)}{\gamma ^2+4 \gamma +11}\,,\,\,\,
	v_{22}=\frac{2(\gamma +1) (\gamma +2)}{\gamma ^2+4 \gamma +11}\,,\,\,\,
	\nonumber
	\\
	&&
	a_1=\frac{2 \left(\gamma ^2+2 \gamma +5\right)}{\gamma ^2+4 \gamma +11}\,,\,\,\,
	\phi_1=-\frac{2 (\gamma +1)}{\gamma ^2+4 \gamma +11}\,.
	\label{fixed_point_exp}
\end{eqnarray}
The above IR geometry is the hyperscaling violation one.
Substituting Eqs. \eqref{fixed_point_sol} and \eqref{fixed_point_exp} into Eq. \eqref{exp_butterfly}, we have
\begin{eqnarray}
	v_B\sim T^{\frac{11+4\gamma+\gamma ^2}{9+2\gamma+\gamma ^2}}.\label{scalingEQ}
\end{eqnarray}
One can see that $v_B\rightarrow 0$ as $T\rightarrow 0$, which can also be proved by numerics.
In addition, from the above equation, the scaling exponent $\alpha$ depends on the model parameter $\gamma$.
\begin{figure}[ht!]
	\centering
	\includegraphics[width=0.46\textwidth]{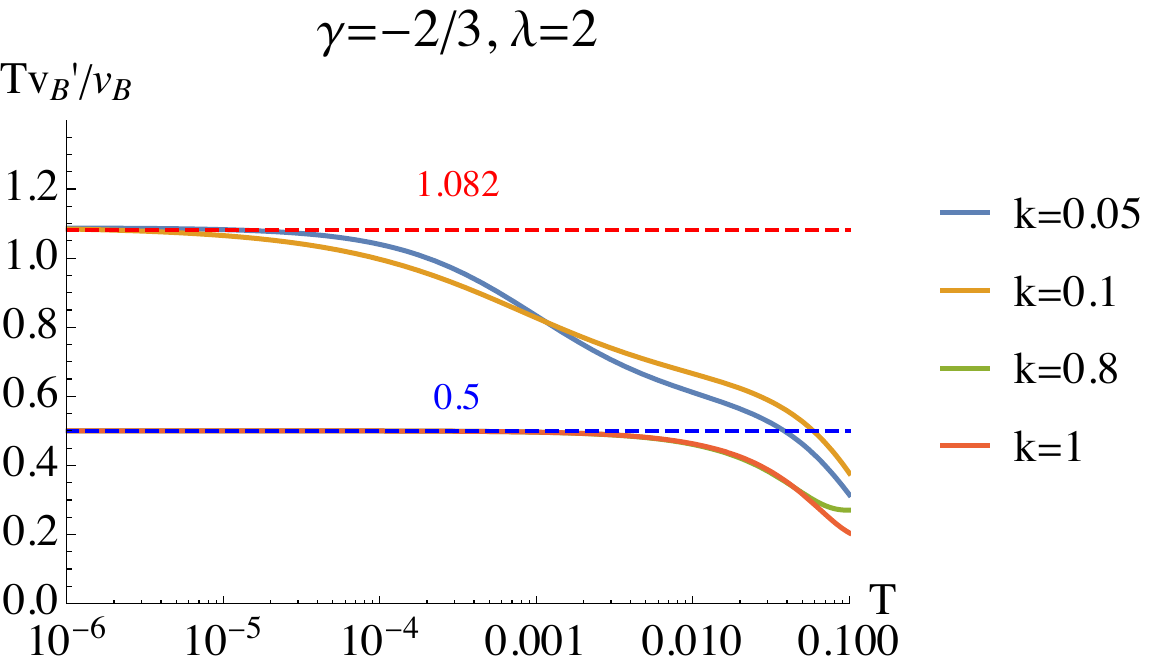}\hspace{5mm}
	\includegraphics[width=0.46\textwidth]{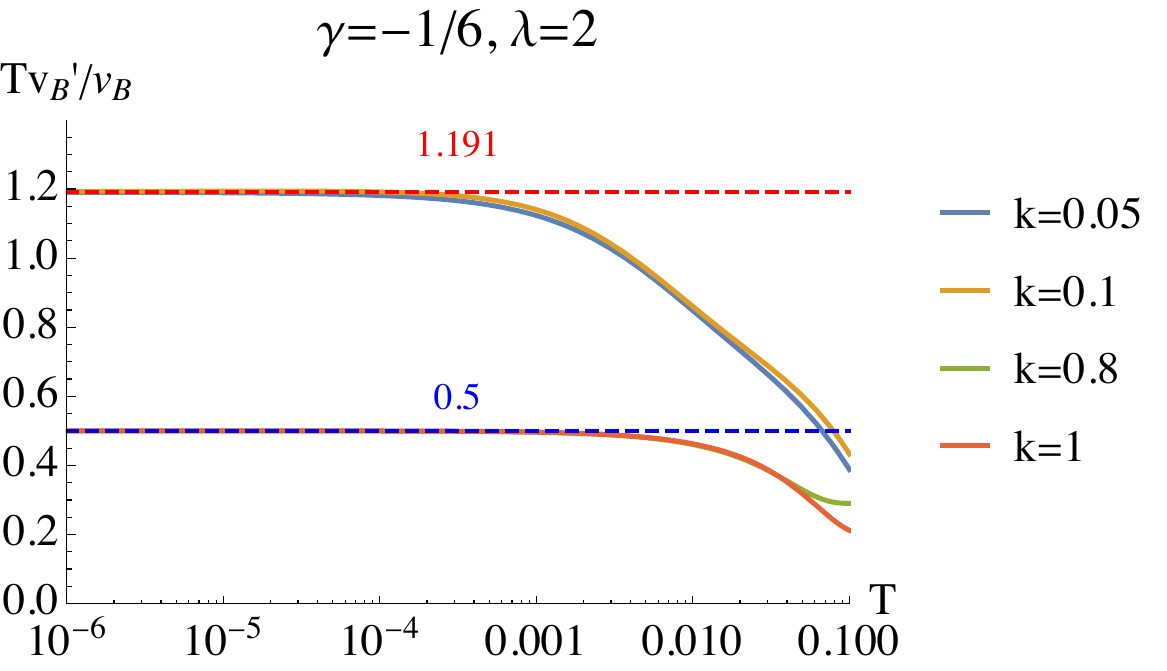}
	\
	\\
	\includegraphics[width=0.46\textwidth]{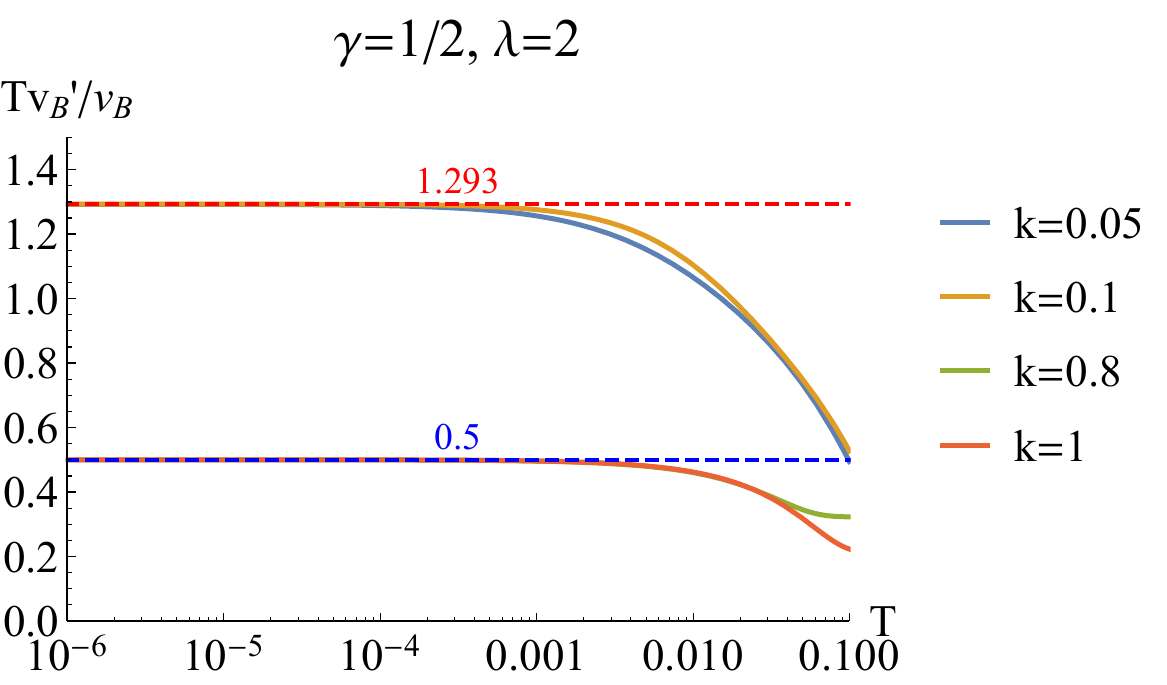}\hspace{5mm}
	\includegraphics[width=0.46\textwidth]{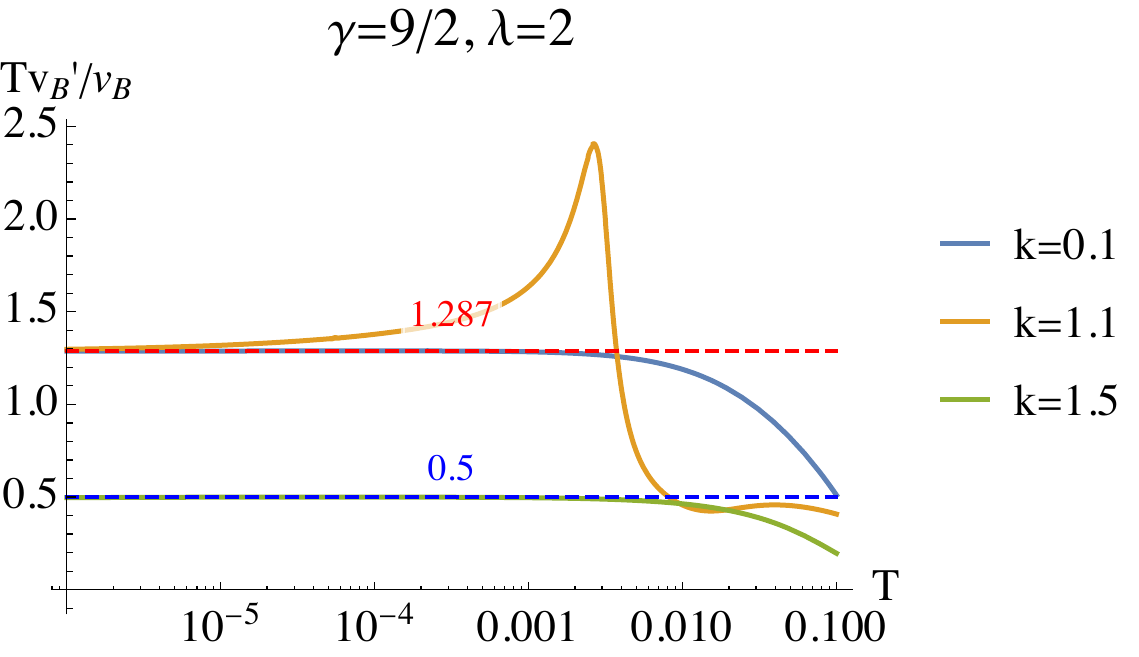}
	\caption{ $T v_B'/v_B$ v.s. $T$. The solid lines indicate the scaling behaviour of $v_B$ with temperature. The blue dashed lines are the scaling exponent of AdS$_2\times \mathbb{R}_2$, for which $\alpha=1/2$. The red dashed lines are the scaling exponent of new IR fixed point, for which $\alpha=\frac{11+4\gamma+\gamma ^2}{9+2\gamma+\gamma ^2}$.}
	\label{FIGscaling}
\end{figure}

Now, we numerically confirm that the theory indeed flows to different IR fixed points for different phases. To this end, we plot $T v_B'/v_B$ versus temperature $T$ with different parameters $\gamma$ (see Fig.\ref{FIGscaling}). The solid lines indicate the scaling behavior of $v_B$ with temperature. From these plots, it is clear to see that in the zero-temperature limit, the theory indeed flows to different IR fixed points for different phases. Furthermore, we would like to point out that for case I, case II and $-1/12<\gamma<3$ of case III, the theory flows to an AdS$_2\times \mathbb{R}_2$ for metallic phase and a hyperscaling violation geometry for insulating phase. The scaling exponent of $v_B$ for AdS$_2\times \mathbb{R}_2$ IR geometry is $\alpha=1/2$ and the one for hyperscaling violation IR geometry is always larger than $1/2$, which depends on the model parameter $\gamma$. However, for $\gamma>3$ of case III, the novel metallic phase flows to a non-AdS$_2\times \mathbb{R}_2$ IR geometry and the normal metallic phase flows to an AdS$_2\times \mathbb{R}_2$ IR geometry. Such a novel phase transition is very different from the previous cases and also different from the result of our previous work on Q-lattice model \cite{Ling:2016ibq}.

\section{Conclusion and discussion}\label{conclusion}

In this paper, we systematically explore the phase structure of the EMDA theory proposed in \cite{Donos:2014uba}. This theory flows to two different IR fixed points: one is the well-known AdS$_2\times \mathbb{R}_2$ and another is the hyperscaling violation geometry, for which the hyperscaling and Lifshitz exponents are determined by the model parameter $\gamma$. QPT happens when there is a transition between different IR fixed points. There are two different mechanisms driving QPT \cite{Donos:2012js}. The usual holographic QPT is induced by the instability of the IR geometry, for example, cases II and III in our EMDA model, and also in the usual Q-lattice model \cite{Donos:2013eha}. Also, QPT is induced by the strength of lattice deformation leading to some kind of bifurcating solution like case I.

The butterfly velocity is an important dynamical quantity to probe the properties of IR geometry. The study on the behavior of the butterfly velocity crossing QCPs indicates that the butterfly velocity $v_B$ or its first derivative exhibiting local extreme depends on the QPT mechanism. If a transition is induced by the instability of the IR geometry, we find that the first derivative of butterfly velocity captures QPT. While if the QPT is induced by the strength of lattice deformation resulting in some kind of bifurcating solution, the QPT is characterized by the butterfly velocity itself. Further, the scaling behaviors of the butterfly velocity in the zero-temperature limit confirm that different phases are controlled by different IR geometries. Therefore, the butterfly velocity is a good probe of QPT and it also provides a possible way to study QPT beyond holography.

Another interesting and important finding is that a novel metallic phase with non-AdS$_2\times \mathbb{R}_2$ IR geometry is found.
For $\gamma>3$, the ground state of this EMDA theory is metallic phase \cite{Donos:2014uba}. The metallic behavior for small $k$ has been confirmed by the DC conductivity and the AC behavior analytically worked out in the zero temperature EMDA geometry. However, this metallic phase has a non-AdS$_2\times \mathbb{R}_2$ IR geometry, which is also confirmed by the scaling behavior of butterfly velocity studied here. With $k$ increasing, there is a phase transition from this novel metallic phase to a normal metallic phase. In this normal metallic phase, the lattice deformation becomes irrelevant in the IR leading to an AdS$_2 \times \mathbb{R}^2$.

Next, we point out some directions worthy of further investigation.
Here, we focused on the phase structure and the butterfly velocity of one-dimensional latticed EMDA model supported by one axion field. It is interesting to extend this model to a two-dimensional latticed system supported by two axion fields, $\chi_1=k_1 x$ and $\chi_2=k_2 y$. In our previous work on two-dimensional Q-latticed model \cite{Liu:2021stu}, we have found the non-AdS$_2 \times \mathbb{R}^2$ IR fixed point for metallic phase. The scaling behavior of the butterfly velocity even does not follow the temperature power-law behavior as $v_B\sim T^\alpha$ \cite{Liu:2021stu}. So it is expected that some novel phenomena appearing in the EMDA model when two axion fields are introduced. We shall explore this issue in near future. 

In this paper, the lattice strength $\lambda$, as the source of the scalar coupling $\psi$, is always turned on. By the UV analysis for the axion field $\chi$, non-zero $\lambda$ leads to the explicit breaking of spatial translation in the dual boundary theory \cite{Amoretti:2017frz,Amoretti:2017axe,Amoretti:2018tzw,Amoretti:2019cef,Amoretti:2019kuf}. If $\lambda= 0$, this may result in spontaneous breaking of the translational symmetry. If the translational symmetry breaking is spontaneous, there should exist gapless excitations in the low energy description called Goldstone modes. Based on the above cases, when turning on the explicit breaking slightly, it will realize the so-called pseudo-spontaneous breaking of translations. And the corresponding excitations are pseudo-Goldstone modes. In this work, the analysis on the shear mode is lacking. It is desirable to study the quasi-normal modes, discuss the phonon dynamics and make a comparison with predictions from hydrodynamics \cite{Amoretti:2019cef,Amoretti:2019kuf,Alberte:2017cch,Alberte:2017oqx,Ammon:2019wci,Ammon:2019apj,Baggioli:2019abx,Baggioli:2020edn,Baggioli:2021xuv,Wang:2021jfu}. It is also interesting to study the diffusion constant in our present holographic framework when the translational symmetry is broken spontaneously, called the crystal diffusion constant or the longitudinal phonon diffusion, which have been explored in the simple axion models \cite{Jeong:2021zhz,Wu:2021mkk,Jeong:2021zsv}.

It is also worthwhile to construct MIT from higher derivative gravities and study the butterfly velocity crossing MIT. There are two main reasons to explore this topic. One is that in the work \cite{Ling:2016dck}, a MIT model was constructed from higher derivative gravity including Weyl tensor, where the second order derivative of holographic entanglement entropy (HEE) with respect to the relevant parameter diagnoses the QPT. It is different from previous works \cite{Ling:2015dma,Ling:2016wyr}, where HEE itself or its first order derivative with respect to the system parameter characterizes the QPT. These results indicate that the QPT could be classified by the order of the derivatives to HEE characterizing the QPT. The other is that the butterfly effect in D-dimensional gravitational theories containing terms quadratic in Ricci scalar and Ricci tensor was studied \cite{Alishahiha:2016cjk}. Due to higher order derivatives in the corresponding equations of motion there are two butterfly velocities. The velocities are determined by the dimension of operators whose sources are provided by the metric. Also they studied the three dimensional topological massive gravity (TMG) \cite{Alishahiha:2016cjk} and found that there are also two butterfly velocities at generic point of the moduli space of parameters. At critical point two velocities coincide.

\acknowledgments

We are very grateful to Zhuo-Yu Xian for helpful discussions and suggestions. This work is supported by the Natural Science Foundation of China under Grant Nos. 11905083, 11775036, 12147209, and Fok Ying Tung Education Foundation
under Grant No. 171006, and the Postgraduate Research \& Practice Innovation Program of Jiangsu Province under Grant No. KYCX20\_2973 and KYCX21\_3192. X. M. Kuang is also supported by Natural Science Foundation of Jiangsu Province under Grant No.BK20211601. J. P. Wu is also supported by Top Talent Support Program from Yangzhou University.

\appendix
\section{Equations of motion}\label{appendix}
In this appendix, we consider a general form of the EMDA action and give the corresponding equations of motion by the variation.
The action we considered is
\begin{eqnarray} \label{appendix_action}
S= {}\int d^{4}x \sqrt{-g} \left[ R -V(\psi)-\frac{c}{2} [ (\partial\psi)^2+Y(\psi)(\partial{\chi})^2 ] - \frac{1}{4}Z(\psi)F^2 \right]\,.
\end{eqnarray}
When $V(\psi)=-6 \cosh\psi$, $Y(\psi)=4\sinh^2\psi$, $Z(\psi)=\cosh^{\gamma /3}(3\psi)$ and $c=3$, the above action reduces to the action \eqref{EMDA-Action} considered in our paper.
From the action \eqref{appendix_action}, we can derive the corresponding equations of motion, which are
\begin{eqnarray} \label{appenix_eom}
&&
R_{\mu\nu}-\frac{1}{2}R g_{\mu\nu}=T_{\mu\nu}^{(V)}+T_{\mu\nu}^{(A)}+T_{\mu\nu}^{(\psi)}+T_{\mu\nu}^{(\chi)} \,,
\
\\
&&
\nabla_\mu(Z F^{\mu\nu})=0\,,
\
\\
&&
\nabla_\mu(Y \nabla^{\mu}\chi)=0\,,
\
\\
&&
c \square\psi-V_{\,, \psi}-\frac{1}{4}Z_{\,, \psi}F^2-\frac{c}{2}Y_{\,, \psi}(\partial\chi)^2=0\,,
\end{eqnarray}
where $T_{\mu\nu}^{\#} $ are the energy-momentum tensor of the matter fields defined as
\begin{eqnarray}
	&&
T_{\mu\nu}^{(V)}=-\frac{1}{2} g_{\mu\nu}V\,,
\
\\
&&
T_{\mu\nu}^{(A)}=-\frac{Z}{2}\left(\frac{1}{4}g_{\mu\nu}F^2-F^{\rho}_{\mu}F_{\rho\nu}\right)\,,
\
\\
&&
T_{\mu\nu}^{(\psi)}=-\frac{c}{4}\left(g_{\mu\nu}\partial\psi^2-2\nabla_\mu\psi\nabla_\nu\psi\right)\,,
\
\\
&&
T_{\mu\nu}^{(\chi)}=-\frac{c}{4}Y\left(g_{\mu\nu}\partial\chi^2-2\nabla_\mu\chi\nabla_\nu\chi\right)\,.
\end{eqnarray}
Without loss of generality, we assume the background geometry to be the following form
\begin{eqnarray}\label{appendix_metric}
	&&
ds^2=-D(r) dt^2+ B(r) dr^2+C_1(r)dx^2+C_2(r)dy^2\,, \nonumber \\
&&
A=a(r)dt\,, \ \ \psi=\psi(r)\,, \ \ \chi=k x\,.
\end{eqnarray}
Using the above ansatz, we can obtain the following equations of motion
\begin{eqnarray} \label{appendix_EOM}
	&&
a''+\left(-\frac{D'}{2D}+\frac{C_1'}{2C_1}+\frac{C_2'}{2C_2}-\frac{B'}{2B}+\frac{Z'}{Z}\right)a'=0\,,
\
\\
&&
\psi''+\left(-\frac{B'}{2B}+\frac{C_1'}{2C_1}+\frac{C_2'}{2C_2}+\frac{D'}{2D}\right)\psi'+\frac{Z_{\,,\psi}}{2 D} a'^2-\frac{k^2B Y_{\,, \psi}}{2C_1}-B  V_{\,, \psi}=0\,,
\
\\
&&
\frac{C_1''}{C_1}+\frac{C_2''}{C_2}-\left(\frac{B'}{2B}+\frac{D'}{2D}\right)\left(\frac{C_1'}{C_1}+\frac{C_2'}{C_2}\right)-\frac{C_1'}{2C_1}^2-\frac{C_2'}{2C_2}^2+c \psi'^2=0 \,,
\
\\
&&
\frac{C_1''}{C_1}+\frac{C_2''}{C_2}-\left(\frac{B'}{2B}-\frac{D'}{2D}\right)\left(\frac{C_1'}{C_1}+\frac{C_2'}{C_2}\right)-\frac{(C_2C_1'-C_1C_2')^2}{2C_1^2C_2^2}+\frac{Z a'^2}{D}+\frac{c k^2 B Y}{C_1}+2 B V=0 \,,
\nonumber
\\
\
\\
&&
\frac{D''}{D}-\frac{D'}{2D}^2-\frac{C_2''}{C_2}+\frac{C_2'}{2C_2}^2-\left(\frac{C_2'}{2C_2}-\frac{D'}{2D}\right)\left(\frac{C_1'}{C_1}-\frac{B'}{B}\right)-\frac{Z a'^2}{D}=0 \,,
\
\\
&&
\frac{D''}{D}-\frac{D'}{2D}^2+\frac{C_2''}{C_2}-\frac{C_2'}{2C_2}^2+\left(\frac{C_2'}{2C_2}+\frac{D'}{2D}\right)\left(\frac{C_1'}{C_1}-\frac{B'}{B}\right)+\frac{C_2' D'}{C_2 D}-2BV=0\,,
\end{eqnarray}
where $\#_{,\psi}$ and the prime denote the derivative with respect to $\psi$ and $r$, respectively. The equation of axion field is automatically satisfied.

\bibliographystyle{style1}
\bibliography{Ref}
\end{document}